\documentclass[fleqn,usenatbib]{mnras}

\usepackage{newtxtext,newtxmath}

\usepackage[T1]{fontenc}
\usepackage{ae,aecompl}

\usepackage{graphicx}	
\usepackage{amsmath}	
\usepackage{amssymb}	
\usepackage{subfig}
\usepackage{xspace}
\usepackage{threeparttable}

\newcommand{\cxo}{{\it Chandra}\xspace}

\def\xmm{\textit{XMM-Newton}\xspace}

\newcommand{\src}{{IKT\,16}\xspace}
\usepackage{xcolor}

\title[ \src: a 22\,ms pulsar in the SMC]{\src a.k.a. PSR\,J0058--7218: Discovery of a 22\,ms energetic rotation-powered pulsar in the Small Magellanic Cloud}

\author[C. Maitra et al.]{C. Maitra,$^{1}$\thanks{E-mail: cmaitra@mpe.mpg.de}
P. Esposito,$^{2,3}$
A. Tiengo,$^{2,3,4}$
J.~Ballet,$^{5}$
F. Haberl,$^{1}$
S. Dai,$^{6,7}$
\newauthor
M. D. Filipovi\'c,$^{6}$
M. Pilia$^{8}$
\\
$^{1}$Max-Planck-Institut f{\"u}r extraterrestrische Physik, Gie{\ss}enbachstra{\ss}e 1, 85748 Garching, Germany\\
$^{2}$Scuola Universitaria Superiore IUSS Pavia, Palazzo del Broletto, piazza della Vittoria 15, 27100 Pavia, Italy\\
$^{3}$INAF--Istituto di Astrofisica Spaziale e Fisica Cosmica di Milano, via A. Corti 12, 20133 Milano, Italy\\
$^{4}$Istituto Nazionale di Fisica Nucleare, Sezione di Pavia, via A. Bassi 6, I–27100 Pavia, Italy \\
$^{5}$AIM, CEA, CNRS, Universit{\'e} Paris-Saclay, Universit{\'e} de Paris, F-91191 Gif sur Yvette, France\\
$^{6}$Western Sydney University, Locked Bag 1797, Penrith, NSW 2751, Australia\\
$^{7}$CSIRO Astronomy and Space Sciences, Australia Telescope National Facility, PO Box 76, Epping, NSW 1710, Australia \\
$^{8}$INAF--Osservatorio Astronomico di Cagliari, Via della Scienza 5, I-09047 Selargius, Italy
}

\date{Accepted XXX. Received YYY; in original form ZZZ}

\pubyear{2020}
\begin{document}
\label{firstpage}
\pagerange{\pageref{firstpage}--\pageref{lastpage}}
\maketitle

\begin{abstract}
 We report here on the discovery with \xmm of pulsations at 22\,ms from the central compact source associated with \src, a supernova remnant in the Small Magellanic Cloud (SMC). 
The measured spin period and spin period derivative correspond to 21.7661076(2)\,ms and $2.9(3)\times10^{-14}$ s\,s$^{-1}$, respectively. Assuming standard spin-down by magnetic dipole radiation, the spin-down power corresponds to $1.1\times10^{38}$\,erg\,s$^{-1}$ implying a Crab-like pulsar. This makes it the most energetic pulsar discovered in the SMC so far and a close analogue of PSR\,J0537--6910, a Crab-like pulsar in the Large Magellanic Cloud. 
The characteristic age of the pulsar is 12\,kyr.
Having for the first time a period measure for this source, we also searched for the signal in archival data collected in radio with the Parkes telescope and in $\gamma$-rays with the \emph{Fermi}/LAT, but no evidence for pulsation was found in these energy bands.
\end{abstract}

\begin{keywords}
ISM: individual objects: \src\ -- ISM: supernova remnants -- Radio continuum: ISM -- Radiation mechanisms: general -- Magellanic Clouds
\end{keywords}
\section{Introduction}
The Large and the Small Magellanic Clouds (LMC and SMC) are gas-rich irregular galaxies orbiting the Milky Way. The relatively close distance of $\sim$60 kpc to the SMC and low Galactic foreground absorption ($N_{\rm H} \sim 6 \times 10^{20}$ cm$^{-2}$) enable the study of its entire X-ray  source population down to a luminosity of $\sim$10$^{33}$ erg s$^{-1}$ \citep{2013A&A...558A...3S}. The recent star formation activity ($\sim$40\,Myr ago), has created an environment where a large number of massive stars are expected, many of which are companions in high mass X-ray binary (HMXB) systems \citep{2004AJ....127.1531H}. In tune with this, a large population of HMXBs (predominantly Be X-ray binaries) has been discovered and extensively studied in the SMC. Pulsations have been detected in 63 of these systems, confirming their nature as neutron stars. These Be X-ray binary pulsars are typically a few 10 million years old and have spin periods ranging from 1 to 2000 s \citep{2016A&A...586A..81H}.  On the other hand, of the `younger' population of isolated neutron stars that constitute the rotation powered pulsars, just a handful in number are known.  Only seven such systems have been discovered in the SMC until now from radio surveys  \citep[][]{2001ApJ...553..367C,1991MNRAS.249..654M,2006ApJ...649..235M,2019MNRAS.487.4332T} and their detection may be prone to several selection effects and observational biases \citep[see for e.g.][]{2020MNRAS.494..500T}. 

Ideal sites to search for young rotation powered pulsars are supernova remnant (SNR) -- pulsar wind nebula (PWN) composites; Composite SNRs are robust indicators of the presence of a young and energetic pulsar powering the PWN by an outflow of relativistic particles that interact with its natal SNR and the surrounding interstellar medium. Only two such systems are known in the SMC at this date, namely DEM\,S5 and \src\, \citep{2019MNRAS.486.2507A,2015A&A...584A..41M}. 
\src\ is a large X-ray and radio-faint SNR, in which a central source of hard X-ray emission was identified using \xmm \citep{2004A&A...421.1031V}. In a detailed analysis using additional \xmm observations, \cite{2011A&A...530A.132O} found substantial evidence that the unresolved point source detected at the center of the SNR is a PWN associated with it. Follow-up observations with \cxo resulted in resolving the PWN at the centre of \src , the first such conclusive evidence in the SMC \citep{2015A&A...584A..41M}.  The putative neutron star at the centre of the PWN could be seen as a point source using \cxo and was $\sim$3 times brighter than the soft, symmetric nebula surrounding it. This pointed to the presence of an energetic pulsar dominated by non-thermal emission. We report here the discovery of pulsations from the central source in \src\ (PSR\,J0058$-$7218 from now), confirming its nature as an energetic rotation powered pulsar. Sect.\,2 presents the X-ray observations and analysis, Sect.\,3 the search for counterparts in the radio and $\gamma$-ray wavebands, and the discussion and the conclusions follow in Sect.\,4.


\section{Observation and analysis}
\label{xray}
\src\ was observed with the European Photon Imaging Camera (EPIC) on board the \xmm\ satellite starting on 2020 March 15 for an orbit (Obsid 0841450101). 
The EPIC-pn \citep{2001A&A...365L..18S} was set in small window mode (timing resolution of 5.6718\,ms), while both MOS detectors \citep{2001A&A...365L..27T} were operating in full frame mode (timing resolution of 2.6\,s). All instruments mounted the medium optical-blocking filter.
The raw data were analyzed with the {\sc XMMSAS} v.\,18.00 software package. 
We searched for periods of high background flaring activity by extracting light curves in the energy range of 7.0--15.0\,keV and removed the time intervals with background rates $\geq$~8 and 2.5\,counts\,ks$^{-1}$~arcmin$^{-2}$ for EPIC-pn and EPIC-MOS, respectively. Finally, the net exposure times for pn, MOS1, MOS2 are 117.8, 124.4 and 125.9\,ks respectively not correcting for instrument deadtime.
The source counts were selected from a circular region with radius of 20\,arcsec. For the EPIC-pn 
camera, the background spectra were accumulated in nearby regions in the same CCD as the source, avoiding as much as possible the  SNR. The total source and background counts for the PN camera are 6546 and 3868 respectively. In case of the EPIC-MOS cameras, which were in full-frame mode, background regions were extracted taking into account the SNR centre and size as measured in \cite{2011A&A...530A.132O}, so to avoid the SNR emission. 


\subsection{X-ray timing analysis}

\begin{figure}
\centering
\resizebox{\hsize}{!}{ \includegraphics[angle=0]{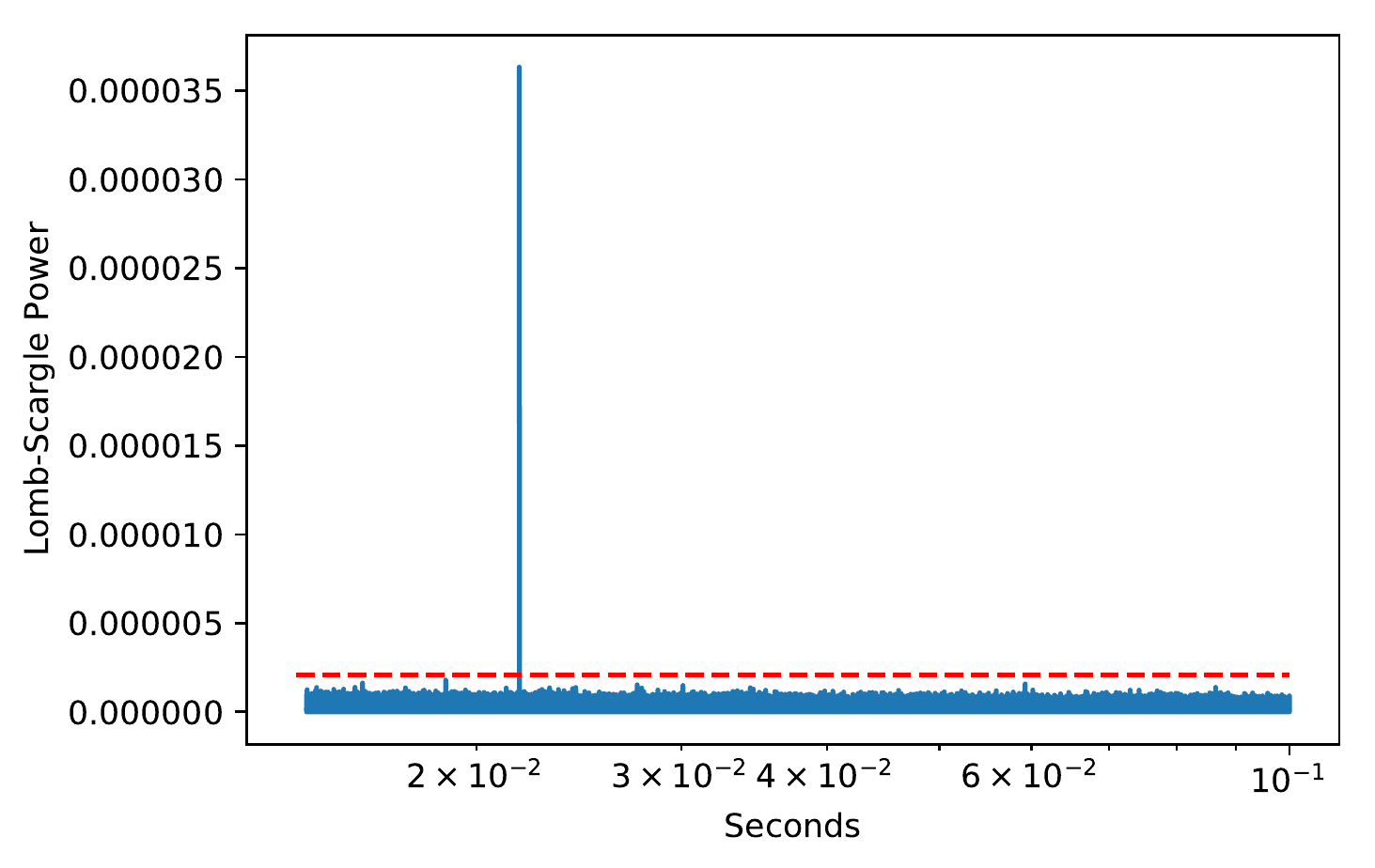}}
   \caption{EPIC-pn Lomb--Scargle periodogram in the energy band 0.4--10\,keV (zoom in). The red dashed line denotes the 3$\sigma$ confidence levels (taking into account the number of periods searched). The highly significant peak at $\sim$22\,ms reflecting the neutron star rotation is apparent.} 
   \label{figtiming}
\end{figure}

To search for a periodic signal, we started with the barycentre-corrected \xmm EPIC-pn data (owing to their better time resolution with respect to the MOS data) in the energy range of 0.4--10.0\,keV by using a Lomb--Scargle periodogram analysis\footnote{https://docs.astropy.org/en/stable/api/astropy.timeseries.LombScargle.html} in the period range of 12\,ms to 1\,s \citep{1976Ap&SS..39..447L,1982ApJ...263..835S}. A strong periodic signal is detected around 22\,ms (Fig.\,\ref{figtiming}). We also verified the result by producing a power density spectra and found a strong peak at the same frequency. The Fourier power with Leahy normalization \citep{1983ApJ...266..160L} corresponded to 592.7 which leads to a chance probability of $3\times10^{-122}$ considering 16777216 trial frequencies searched. A similar exercise performed by extracting events from a large background region did not produce the same peak, thus firmly establishing its origin from the source.

In order to determine the period more precisely, we employed the Bayesian periodic signal detection method described by \cite{1996ApJ...473.1059G}.
 The spin period and its associated 1$\sigma$ error are determined to  21.7661095(1)\,ms (without taking into account a period derivative) indicating the spin period of the neutron star in the centre of \src\,.

\begin{figure}
\centering
\includegraphics[angle=0,scale=0.85]{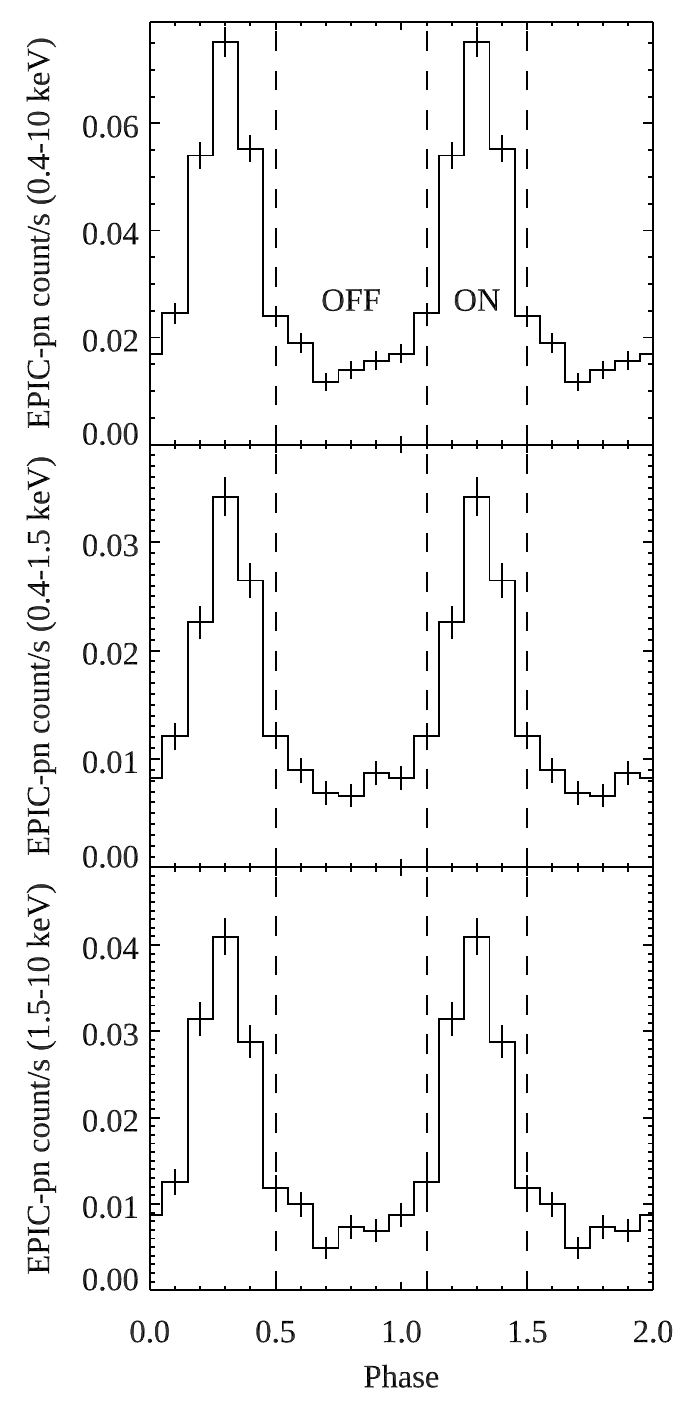}
\caption{Background-subtracted EPIC-pn light curve folded with the time solution of Table\,\ref{timing} in the 0.4--10\,keV (top panel), 0.4--1.5\,keV (middle panel) and 1.5--10\,keV (bottom panel) energy bands. The dashed vertical lines indicate the on-pulse and off-pulse phase intervals used to study the phase-resolved spectrum.}
\label{figpp}
\end{figure}

To check for a period derivative, we divided the exposure in 9 segments and performed a phase-fitting analysis \citep[see e.g.][]{1992RSPTA.341...39P}. A  constant period (a first-order polynomial function) is clearly incompatible with the data, with a reduced $\chi^2$ ($\chi^2_\nu$) of 12.85 for 7 degrees of freedom (dof). The introduction of a period derivative (a second-order polynomial) yields an acceptable fit ($\chi^2_\nu=1.26$ for 6 dof). The best-fitting period derivative is $\dot{P}= 2.9(3)\times10^{-14}$\,s\,s$^{-1}$. We give the complete phase-coherent spin ephemeris and the derived pulsar properties in Table\,\ref{timing}, and in Fig.\,\ref{figpp} we show the epoch folded pulse profiles in different energy bands. The pulse profiles are single peaked in all energy bands and do not show any apparent energy dependence. 
The pulsed fraction, 
defined as $(M-m)/(M+m)$, where $M$ is the maximum of the pulse profile and $m$ the minimum, is 
$(73\pm 3)\%$, $(68\pm 5)\%$ and $(79\pm 5)\%$ 
in the 0.4--10\,keV, 0.4--1.5\,keV and 1.5--10\,keV  energy bands, respectively. As explained in the next session, the pulsar X-ray emission is contaminated, mainly in the soft energy band, by its PWN and by the surrounding SNR and, therefore, these values should be interpreted as lower limits to the pulsar intrinsic pulsed fraction. 


\begin{table} 
\centering
\caption{Spin ephemeris of PSR\,J0058--7218. We also give for convenience the corresponding period $P$ and period derivative $\dot{P}$, as well as the derived characteristic age $\tau_{\mathrm{c}}=P/(2\dot{P})$, dipolar magnetic field $B\approx(3c^3IP\dot P/(8\pi^2R^6))^{1/2}$, and rotational energy loss $\dot{E}=4\pi^2 I \dot{P}P^{-3}$ (here we took $R=10$\,km and $I=10^{45}\ {\rm g\, cm}^2$ for the star radius and moment of inertia, respectively). Figures in parentheses represent the 1$\sigma$ uncertainties in the least significant digit.} 
\begin{tabular}{lc} 
\hline
Parameter      & Value       \\ 
\hline 
Range (MJD) & 58924.016--58925.583\\
Epoch (MJD) & 58924.0\\
Frequency, $\nu$ (Hz) & 45.9429870(4)\\
Frequency derivative, $\dot{\nu}$ (Hz\,s$^{-1}$) & --$6.1(6)\times10^{-11}$\\
Period, $P$ (ms) & 21.7661076(2)\\
Period derivative, $\dot{P}$ (s\,s$^{-1}$) & $2.9(3)\times10^{-14}$\\
Characteristic age, $\tau_{\mathrm{c}}$, (kyr) & 12 \\ 
Spin down luminosity, $\dot{E}$ (erg\,s$^{-1}$) & $1.1\times10^{38}$  \\
Surface dipole magnetic field, $B_{\mathrm{surf}}$ (G) & $8\times10^{11}$ \\ 
\hline 
\label{timing} 
\end{tabular} 
\end{table}


\subsection{X-ray spectral analysis}
Due to the small size and non-uniformity of the SNR and PWN associated to PSR\,J0058--7218, these diffuse components cannot be safely subtracted from the pulsar \xmm\ spectrum selecting a nearby background region. However, since the SNR spectrum is rather different from that of the central point source, 
we defined a background region away from the SNR shock boundary as defined in \cite{2011A&A...530A.132O}, and modelled the central source and the SNR emission components in \src simultaneously in the spectral fit. 
Since the EPIC-pn was operated in small window mode, 
such a background region could be properly defined away from the SNR region only for the MOS cameras. On the other hand, the time resolution of the pn data and the discovery of pulsations allow us to analyse, for the first time, the spectrum of the pulsed emission. In this case, the contribution from both the contaminating sources can be completely eliminated by subtracting the off-pulse emission from the on-pulse spectrum extracted from the same region. The corresponding phase intervals are indicated in Fig.\,\ref{figpp}.  

For the spectral analysis, the XMMSAS tasks \texttt{rmfgen} and \texttt{arfgen} were used to create the redistribution matrices and ancillary files. Spectra were binned to achieve a minimum of 25 counts per spectral bin.
The spectral analysis was performed using the {\small XSPEC} fitting package
\citep{1996ASPC..101...17A}. The X-ray absorption was modeled using the {\texttt tbabs} model \citep{2000ApJ...542..914W} with atomic cross sections adopted from \cite{1996ApJ...465..487V}. We used two absorption components: The first one describes the Galactic foreground absorption, where we used a fixed column density of $6\times 10^{20}$~cm$^{-2}$ \citep{1990ARA&A..28..215D} with abundances taken from \cite{2000ApJ...542..914W}. The second component accounts for the unknown SMC material in front of the object. For the latter absorption component, the abundances were set to SMC abundances (20\% of the solar abundances). 
We adopted a power-law model to 
account for the emission from the pulsar. 
In the MOS phase-averaged spectrum, we included a Sedov component (\texttt{vsedov}) 
to model the SNR emission.
The derived $N_{\rm H}$ and  SNR parameters are consistent with those obtained in \cite{2011A&A...530A.132O}. The power-law index 
is $\Gamma\sim1.4$ both in the total and in the pulsed spectrum. 
Since the PWN emission has a softer spectrum \citep[][]{2015A&A...584A..41M}, this indicates that, although it cannot be spatially resolved using \xmm, its contribution to the total spectrum is marginal.
The unabsorbed phase-averaged luminosity of PSR\,J0058$-$7218 in the 0.2--12\,keV band is $\sim$10$^{35}$ erg~s$^{-1}$  (see Table \ref{spec-tab}). The best-fit parameters of the total and pulsed spectra of PSR\,J0058$-$7218 and the errors corresponding to the 90\% confidence range are tabulated in Table\,\ref{spec-tab}. 
\begin{figure}
\hspace{-0.3cm}
\includegraphics[scale=0.35]{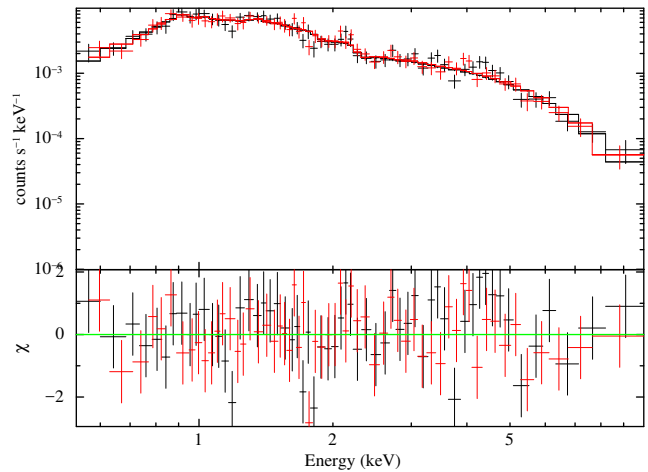}
\caption{\xmm\ EPIC MOS spectra of  PSR\,J0058–-7218 showing MOS1 in black and MOS2 in red. The dotted lines show the contributions of the thermal emission from the SNR (soft) and the power-law emission from the pulsar and the nebula (hard). The residuals for the best-fit model are shown in the lower panel. }
\label{figspec}
\end{figure}

%
\begin{table}
\caption{Best-fit parameters of the phase-averaged (EPIC MOS) and pulsed (EPIC pn) X-ray spectrum of PSR\,J0058–-7218. The line-of-sight Galactic absorption was fixed to 6 $\times$ $10^{20}$~cm$^{-2}$ and the average spectrum was fitted simultaneously with the SNR emission. Errors are quoted at 90\% confidence.}
\begin{tabular}{lcc}
 \hline
   Parameter & Total emission & Pulsed emission$^{b}$ \\
 \hline
   $N_{\mathrm{H}} ^{{\rm local}}$  (10$^{21}$ cm$^{-2}$) & $4\pm2$ & $6\pm2$\\
   $\Gamma$ & $1.4\pm0.1$ & $1.4\pm0.1$\\
   Shock $kT$ (KeV) &  $1.0\pm0.5$ & -- \\
   Ionization timescale (10$^{11}$\,cm$^{-3}$s) & $1.3\pm0.5$ & -- \\
   X-ray luminosity$^{a}$ ($10^{35}$ erg~s$^{-1}$) & $1.2\pm0.1$ & $1.6\pm0.1$\\
   \hline
    $\chi^{2}$       & 100 & 177\\
   Degrees of Freedom & 111 & 156   \\
   \hline
\end{tabular}

$^{a}$Unabsorbed luminosity in the energy band of 0.2--12\,keV, assuming a distance of 60\,kpc. 
$^{b}$ Pulsed flux is evaluated with respect to 40\% of the duty cycle, see Fig.~\ref{figpp}.

\label{spec-tab}
\end{table}

\section{Search for radio and $\gamma$-ray counterpart sources}
\subsection{Search for a radio counterpart}
A 4.5-hour long observation pointing at $\rm RA=00^h58^m06\fs931$, $\rm Decl.=-72^\circ19'17\farcs616$, $\sim$2.1\,arcmin from the pulsar, was taken with the Parkes telescope on 2017 August 28 under the project P944 \citep{2019MNRAS.487.4332T}.
The data were recorded at a central frequency of 1382\,MHz with a bandwidth of 400\,MHz and 1024 frequency channels. The sampling time is $64\, \upmu$s and only the total intensity was recorded with 2-bit sampling. According to the free electron density model YNW16 \citep{yao2017}, the dispersion measure (DM) towards the source is expected to be of the order of 140\,pc\,cm$^{-3}$. Another estimate for DM can be derived by the X-ray absorption column density, $N_{\rm H}$, assuming an ionization fraction. In the Milky Way, the average ionization is $\sim$10\%, as determined empirically from X-ray and radio measurements \citep{hnk13}.
The relation is not calibrated for the Magellanic Clouds, but if we assume 10\%, the expected DM is consistently $\sim$150\,pc\,cm$^{-3}$.

We performed a first targeted search by folding the Parkes data with the 
position and spin parameters obtained from the X-ray observation. The {\tt DSPSR} software package \citep{vanstraten11} was used to fold the data with a sub-integration length of 10\,s. The {\sc PDMP} tool as a part of the {\tt PSRCHIVE} software package \citep{hotan04} was then used to search through a range of barycentric period and DM (0 to 1000\,pc\,cm$^{-3}$) to find values giving the highest signal-to-noise ratio for potential pulsar signals. We also carried out a blind search, but no periodic signal was found over a DM range of 0 to 500\,pc\,cm$^{-3}$ using the {\tt PRESTO} software package\footnote{\url{https://www.cv.nrao.edu/~sransom/presto/}} \citep{ransom01,ransom02}.
No obvious detection was found in these searches, down to the data set sensitivity of 15\,$\upmu$Jy as calculated using the modified radiometer formula for pulsars \citep{lk04}. We also performed a search for single radio pulses using the Python-based pipeline named {\tt SPANDAK}\footnote{\url{https://github.com/gajjarv/PulsarSearch}} from \cite{gajjar18}. Data were first processed through {\tt rfifind} from the {\tt PRESTO} package for high-level radio frequency interferences purging.
The pipeline is based on {\tt Heimdall} \citep{barsdell12} as its main kernel and we used it to search across a DM range from 0 to 1000 pc~cm$^{-3}$. The de-dispersed time-series were searched for pulses using a matched-filtering technique with a minimum window size equalling our time resolution (64~$\upmu$s) and a maximum window size of 65 ms. The pipeline produced around 10 candidates at different DMs. Each candidate found by {\tt Heimdall} was scrutinized to identify possible bursts but none were found down to a limiting fluence of 160\,mJy ms. 

\subsection{Search for a $\gamma$-ray counterpart}
A young pulsar with such a large $\dot{E}$ is expected to emit strongly in $\gamma$ rays and could be visible for $Fermi$-LAT, if the viewing angle is favourable. For example, PSR\,J0540$-$6919 in the Large Magellanic Cloud is easily detected at $2.8 \pm 0.2 \times 10^{-11}$\,erg\,cm$^{-2}$\,s$^{-1}$ (0.1--100 GeV). We have looked for a possible $\gamma$-ray counterpart in the 4FGL-DR2 catalogue \citep{4FGL,4FGL-DR2}. A faint source (4FGL\,J0059.7$-$7210 at $1.6 \pm 0.4 \times 10^{-12}$\,erg\,cm$^{-2}$\,s$^{-1}$) is present 0.18$^\circ$ away from the pulsar's position. The pulsar is just outside the 95\% error ellipse of the LAT source, so it could be the counterpart. However, the 4FGL-DR2 catalogue lists the brightest star-forming region in the SMC (NGC\,346) as a plausible counterpart. It is closer to the LAT source than the pulsar, so we view the LAT flux as an upper limit to the pulsar's $\gamma$-ray flux. The $\gamma$-ray spectral shape (curved, peaking just below 1 GeV) can be explained by either a pulsar or a star-forming region.

The LAT photons were folded at the frequency and frequency derivative obtained from the \xmm\ observation, taking the source position from $Chandra$, and no $\gamma$-ray pulsation was seen (D.~Smith, private communication). Such a negative result does not make it possible to place a constraining upper limit to the pulsed flux, however, because the LAT source (if it is the pulsar) detects only $\sim$40 photons per year, and maintaining phase coherence over 10 years for a young pulsar requires a more complex ephemeris.
\section{Discussion and conclusions}
We report the discovery of a young and energetic pulsar PSR\,J0058--7218 inside the SNR \src\ in the SMC. X-ray pulsations are detected corresponding to a period of 21.7661076(2)\,ms. The pulsar spins down at a rate of $2.9(3)\times10^{-14}$ s\,s$^{-1}$ consistent with the scenario expected from a young rotation powered pulsar. Neither radio nor $\gamma$-ray pulsations are detected from the existing archival observations using the Parkes and $Fermi$/LAT data respectively. We further extracted the pulsed X-ray emission from PSR\,J0058--7218 in order to constrain the pulsar's spectral shape and luminosity which is otherwise contaminated by the underlying PWN and SNR emission in the phase averaged spectrum. A photon-index of 1.4 is consistent with that measured typically in other young rotation-powered pulsars where the non-thermal radiation is generated by the particles accelerated in the pulsar magnetosphere \citep[see][]{1997A&A...326..682B,2004ApJ...601.1038W}. 

The X-ray pulse profile of PSR\,J0058--7218 is single peaked with a duty-cycle of $\simeq$0.4, and shows no apparent evolution with energy in the range of 0.4--10\,keV. The measured spin-period and the spin period derivative can be used to derive the spin down luminosity ($\dot{E}$), characteristic age ($\tau_{\mathrm{c}}$) and the equatorial surface magnetic field ($B_{\mathrm{surf}}$) of the pulsar, assuming spin-down by magnetic dipole radiation. The derived parameters of PSR\,J0058--7218 are given in Table\,\ref{timing}. A spin down luminosity of $\dot{E} \gtrsim$ $10^{38}$ erg\,s$^{-1}$ indicates a Crab-like pulsar making it the first such discovery in the SMC and the third in the Magellanic Clouds, the two others being PSR\,J0537--6910 and PSR\,J0540--6919 located in the 30 Doradus region of the LMC.  A characteristic age of 12\,kyr makes it the oldest known Crab-like pulsar. The estimated age is however in line with the Sedov age of 14.7\,kyr, derived from the X-ray emission of the SNR \citep{2011A&A...530A.132O}. 

\begin{figure}
\centering
\includegraphics[scale=0.35]{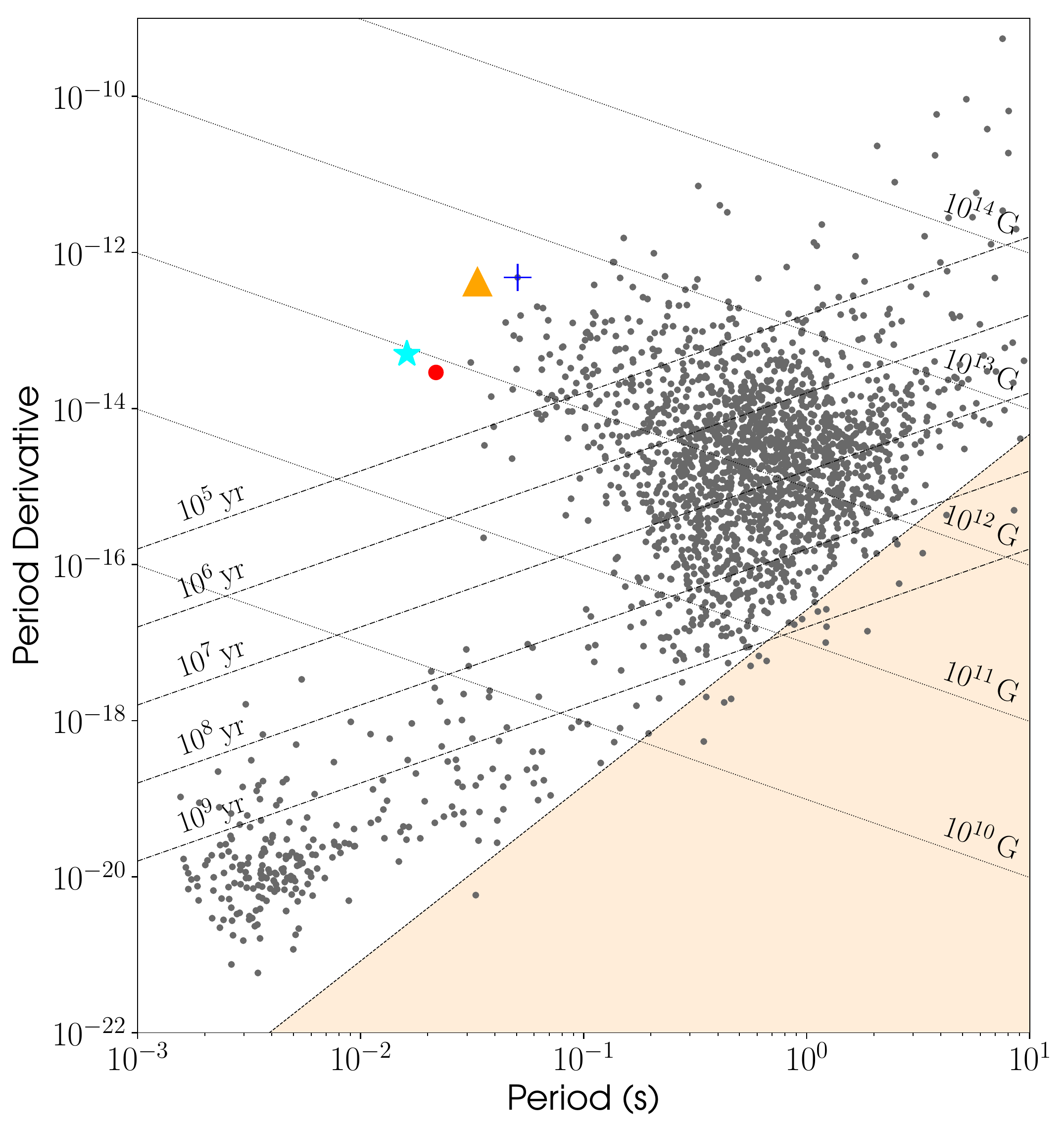}
\caption{P-Pdot diagram showing the position of PSR\,J0058--7218 (red dot), PSR\,J0537--6910 (cyan star), PSR\,J0540--6919 (orange triangle) and the Crab pulsar (blue cross).}
\label{figppdot}
\end{figure}

Figure\,\ref{figppdot} shows the location of PSR\,J0058--7218 on the $P$--$\dot{P}$ diagram of pulsars, making it a close analogue of PSR\,J0537--6910 \citep[the fastest rotating non-recycled pulsar in the LMC SNR N157B;][]{1998ApJ...499L.179M} in terms of their spin-down properties. The radiation
efficiency $\eta_{\rm x}$, which is defined as the ratio of the pulsar's non-thermal X-ray radiation luminosity to the spin-down luminosity \citep[see e.g.][]{2016ApJ...833...59S}, is also comparable for the two systems and is of the order of $\sim$10$^{-3}$. However, this is about two orders of magnitude smaller than that of the Crab pulsar and PSR\, J0540-6919. The observed differences (and similarities) in radiation efficiencies can be accounted for by a geometric effect mainly due to differences in the observer's viewing angle \citep{2017ApJ...834....4T}. 
PSR\,J0537--6910 is also a radio-quiet pulsar, emitting non-thermal X-ray pulsed emission and without detected $\gamma$-ray pulsations. This makes it similar to our system at the current detection limits of the radio and $\gamma$-ray observations. The above observed properties could also indicate a geometric effect as the non-detection of radio emission and the thermal X-ray component
imply that the polar cap region is hidden from our view.
 In this case, as the beaming directions of the $\gamma$-rays and the non-thermal X-rays may not be the same, our line of sight might also miss the $\gamma$-ray beam \citep{1998ApJ...493L..35C,2017ApJ...834..120H}. 
 
  The discovery of a young, energetic and ultra-fast pulsar like PSR\,J0058--7218 provides a unique opportunity to probe the braking mechanisms and birth-spin models of rotation-powered pulsars. Future monitoring of PSR\,J0058--7218 is crucial to constrain the second derivative of the period in order to measure the braking-index of the pulsar and allow deeper searches in the radio and $\gamma$-rays, and look for putative glitches that are fairly common in young rotation powered pulsars on timescales of a few years. A continuous monitoring of the spin evolution will also be very important because of its potential as a source of detectable gravitational waves \citep{2019ApJ...879...10A}.

\section*{Acknowledgements}
The authors thank the referee for useful comments and suggestions. This work uses observations obtained with \xmm, an ESA science mission with instruments and contributions directly funded by ESA Member States and NASA. The \xmm project is supported by the DLR and the Max Planck Society. PE and AT acknowledge funding from PRIN/MIUR award 2017LJ39LM. JB acknowledges support by the Centre National d'Etudes Spatiales (CNES). 
\section*{Data Availability}

X-ray data are available through the High Energy Astrophysics Science Archive Research Center \url{heasarc.gsfc.nasa.gov}. 
The radio data are publicly available from \url{https://data.csiro.au/}.C



\bibliographystyle{mnras}
\bibliography{example,biblioRadio} 

\begin{thebibliography}{}
\makeatletter
\relax
\def\mn@urlcharsother{\let\do\@makeother \do\$\do\&\do\#\do\^\do\_\do\%\do\~}
\def\mn@doi{\begingroup\mn@urlcharsother \@ifnextchar [ {\mn@doi@}
  {\mn@doi@[]}}
\def\mn@doi@[#1]#2{\def\@tempa{#1}\ifx\@tempa\@empty \href
  {http://dx.doi.org/#2} {doi:#2}\else \href {http://dx.doi.org/#2} {#1}\fi
  \endgroup}
\def\mn@eprint#1#2{\mn@eprint@#1:#2::\@nil}
\def\mn@eprint@arXiv#1{\href {http://arxiv.org/abs/#1} {{\tt arXiv:#1}}}
\def\mn@eprint@dblp#1{\href {http://dblp.uni-trier.de/rec/bibtex/#1.xml}
  {dblp:#1}}
\def\mn@eprint@#1:#2:#3:#4\@nil{\def\@tempa {#1}\def\@tempb {#2}\def\@tempc
  {#3}\ifx \@tempc \@empty \let \@tempc \@tempb \let \@tempb \@tempa \fi \ifx
  \@tempb \@empty \def\@tempb {arXiv}\fi \@ifundefined
  {mn@eprint@\@tempb}{\@tempb:\@tempc}{\expandafter \expandafter \csname
  mn@eprint@\@tempb\endcsname \expandafter{\@tempc}}}

\bibitem[\protect\citeauthoryear{{Abbott} et~al.,}{{Abbott}
  et~al.}{2019}]{2019ApJ...879...10A}
{Abbott} B.~P.,  et~al., 2019, \mn@doi [\apj] {10.3847/1538-4357/ab20cb}, \href
  {https://ui.adsabs.harvard.edu/abs/2019ApJ...879...10A} {879, 10}

\bibitem[\protect\citeauthoryear{{Abdollahi}, {Acero}, {Ackermann}, {Ajello},
  {Atwood}  et~al.}{{Abdollahi} et~al.}{2020}]{4FGL}
{Abdollahi} S.,  {Acero} F.,  {Ackermann} M.,  {Ajello} M.,  {Atwood} W.~B.,
  et~al., 2020, \mn@doi [\apjs] {10.3847/1538-4365/ab6bcb}, \href
  {https://ui.adsabs.harvard.edu/abs/2020ApJS..247...33A} {247, 33}

\bibitem[\protect\citeauthoryear{{Alsaberi} et~al.,}{{Alsaberi}
  et~al.}{2019}]{2019MNRAS.486.2507A}
{Alsaberi} R. Z.~E.,  et~al., 2019, \mn@doi [\mnras] {10.1093/mnras/stz971},
  \href {https://ui.adsabs.harvard.edu/abs/2019MNRAS.486.2507A} {486, 2507}

\bibitem[\protect\citeauthoryear{{Arnaud}}{{Arnaud}}{1996}]{1996ASPC..101...17A}
{Arnaud} K.~A.,  1996, in {Jacoby} G.~H.,  {Barnes} J.,  eds,  Astronomical
  Society of the Pacific Conference Series Vol. 101, Astronomical Data Analysis
  Software and Systems V. p.~17

\bibitem[\protect\citeauthoryear{{Ballet}, {Burnett}, {Digel}  \&
  {Lott}}{{Ballet} et~al.}{2020}]{4FGL-DR2}
{Ballet} J.,  {Burnett} T.~H.,  {Digel} S.~W.,   {Lott} B.,  2020, arXiv
  e-prints, \href {https://ui.adsabs.harvard.edu/abs/2020arXiv200511208B} {p.
  arXiv:2005.11208}

\bibitem[\protect\citeauthoryear{{Barsdell}, {Bailes}, {Barnes}  \&
  {Fluke}}{{Barsdell} et~al.}{2012}]{barsdell12}
{Barsdell} B.~R.,  {Bailes} M.,  {Barnes} D.~G.,   {Fluke} C.~J.,  2012,
  \mn@doi [MNRAS] {10.1111/j.1365-2966.2012.20622.x}, \href
  {http://adsabs.harvard.edu/abs/2012MNRAS.422..379B} {422, 379}

\bibitem[\protect\citeauthoryear{{Becker} \& {Truemper}}{{Becker} \&
  {Truemper}}{1997}]{1997A&A...326..682B}
{Becker} W.,  {Truemper} J.,  1997, \aap, \href
  {https://ui.adsabs.harvard.edu/abs/1997A&A...326..682B} {326, 682}

\bibitem[\protect\citeauthoryear{{Cheng}, {Gil}  \& {Zhang}}{{Cheng}
  et~al.}{1998}]{1998ApJ...493L..35C}
{Cheng} K.~S.,  {Gil} J.,   {Zhang} L.,  1998, \mn@doi [\apjl]
  {10.1086/311117}, \href
  {https://ui.adsabs.harvard.edu/abs/1998ApJ...493L..35C} {493, L35}

\bibitem[\protect\citeauthoryear{{Crawford}, {Kaspi}, {Manchester}, {Lyne},
  {Camilo}  \& {D'Amico}}{{Crawford} et~al.}{2001}]{2001ApJ...553..367C}
{Crawford} F.,  {Kaspi} V.~M.,  {Manchester} R.~N.,  {Lyne} A.~G.,  {Camilo}
  F.,   {D'Amico} N.,  2001, \mn@doi [\apj] {10.1086/320635}, \href
  {https://ui.adsabs.harvard.edu/abs/2001ApJ...553..367C} {553, 367}

\bibitem[\protect\citeauthoryear{{Dickey} \& {Lockman}}{{Dickey} \&
  {Lockman}}{1990}]{1990ARA&A..28..215D}
{Dickey} J.~M.,  {Lockman} F.~J.,  1990, \mn@doi [\araa]
  {10.1146/annurev.aa.28.090190.001243}, \href
  {http://adsabs.harvard.edu/abs/1990ARA%26A..28..215D} {28, 215}

\bibitem[\protect\citeauthoryear{{Gajjar} et~al.,}{{Gajjar}
  et~al.}{2018}]{gajjar18}
{Gajjar} V.,  et~al., 2018, \mn@doi [\apj] {10.3847/1538-4357/aad005}, \href
  {http://adsabs.harvard.edu/abs/2018ApJ...863....2G} {863, 2}

\bibitem[\protect\citeauthoryear{{Gregory} \& {Loredo}}{{Gregory} \&
  {Loredo}}{1996}]{1996ApJ...473.1059G}
{Gregory} P.~C.,  {Loredo} T.~J.,  1996, \mn@doi [\apj] {10.1086/178215}, \href
  {https://ui.adsabs.harvard.edu/abs/1996ApJ...473.1059G} {473, 1059}

\bibitem[\protect\citeauthoryear{{Haberl} \& {Sturm}}{{Haberl} \&
  {Sturm}}{2016}]{2016A&A...586A..81H}
{Haberl} F.,  {Sturm} R.,  2016, \mn@doi [\aap] {10.1051/0004-6361/201527326},
  \href {http://adsabs.harvard.edu/abs/2016A%26A...586A..81H} {586, A81}

\bibitem[\protect\citeauthoryear{{Harris} \& {Zaritsky}}{{Harris} \&
  {Zaritsky}}{2004}]{2004AJ....127.1531H}
{Harris} J.,  {Zaritsky} D.,  2004, \mn@doi [\aj] {10.1086/381953}, \href
  {https://ui.adsabs.harvard.edu/abs/2004AJ....127.1531H} {127, 1531}

\bibitem[\protect\citeauthoryear{{He}, {Ng}  \& {Kaspi}}{{He}
  et~al.}{2013}]{hnk13}
{He} C.,  {Ng} C.~Y.,   {Kaspi} V.~M.,  2013, \mn@doi [\apj]
  {10.1088/0004-637X/768/1/64}, \href
  {https://ui.adsabs.harvard.edu/abs/2013ApJ...768...64H} {768, 64}

\bibitem[\protect\citeauthoryear{{Hotan}, {van Straten}  \&
  {Manchester}}{{Hotan} et~al.}{2004}]{hotan04}
{Hotan} A.~W.,  {van Straten} W.,   {Manchester} R.~N.,  2004, \mn@doi [\pasa]
  {10.1071/AS04022}, \href
  {https://ui.adsabs.harvard.edu/abs/2004PASA...21..302H} {21, 302}

\bibitem[\protect\citeauthoryear{{Hui}, {Lee}, {Takata}, {Ng}  \&
  {Cheng}}{{Hui} et~al.}{2017}]{2017ApJ...834..120H}
{Hui} C.~Y.,  {Lee} J.,  {Takata} J.,  {Ng} C.~W.,   {Cheng} K.~S.,  2017,
  \mn@doi [\apj] {10.3847/1538-4357/834/2/120}, \href
  {https://ui.adsabs.harvard.edu/abs/2017ApJ...834..120H} {834, 120}

\bibitem[\protect\citeauthoryear{{Leahy}, {Darbro}, {Elsner}, {Weisskopf},
  {Sutherland}, {Kahn}  \& {Grindlay}}{{Leahy}
  et~al.}{1983}]{1983ApJ...266..160L}
{Leahy} D.~A.,  {Darbro} W.,  {Elsner} R.~F.,  {Weisskopf} M.~C.,  {Sutherland}
  P.~G.,  {Kahn} S.,   {Grindlay} J.~E.,  1983, \mn@doi [\apj]
  {10.1086/160766}, \href
  {https://ui.adsabs.harvard.edu/abs/1983ApJ...266..160L} {266, 160}

\bibitem[\protect\citeauthoryear{{Lomb}}{{Lomb}}{1976}]{1976Ap&SS..39..447L}
{Lomb} N.~R.,  1976, \mn@doi [\apss] {10.1007/BF00648343}, \href
  {https://ui.adsabs.harvard.edu/abs/1976Ap&SS..39..447L} {39, 447}

\bibitem[\protect\citeauthoryear{{Lorimer} \& {Kramer}}{{Lorimer} \&
  {Kramer}}{2004}]{lk04}
{Lorimer} D.~R.,  {Kramer} M.,  2004, {Handbook of Pulsar Astronomy}.
Cambridge, UK: Cambridge University Press

\bibitem[\protect\citeauthoryear{{Maitra}, {Ballet}, {Filipovi{\'c}}, {Haberl},
  {Tiengo}, {Grieve}  \& {Roper}}{{Maitra} et~al.}{2015}]{2015A&A...584A..41M}
{Maitra} C.,  {Ballet} J.,  {Filipovi{\'c}} M.~D.,  {Haberl} F.,  {Tiengo} A.,
  {Grieve} K.,   {Roper} Q.,  2015, \mn@doi [\aap]
  {10.1051/0004-6361/201526458}, \href
  {https://ui.adsabs.harvard.edu/abs/2015A&A...584A..41M} {584, A41}

\bibitem[\protect\citeauthoryear{{Manchester}, {Fan}, {Lyne}, {Kaspi}  \&
  {Crawford}}{{Manchester} et~al.}{2006}]{2006ApJ...649..235M}
{Manchester} R.~N.,  {Fan} G.,  {Lyne} A.~G.,  {Kaspi} V.~M.,   {Crawford} F.,
  2006, \mn@doi [\apj] {10.1086/505461}, \href
  {https://ui.adsabs.harvard.edu/abs/2006ApJ...649..235M} {649, 235}

\bibitem[\protect\citeauthoryear{{Marshall}, {Gotthelf}, {Zhang}, {Middleditch}
   \& {Wang}}{{Marshall} et~al.}{1998}]{1998ApJ...499L.179M}
{Marshall} F.~E.,  {Gotthelf} E.~V.,  {Zhang} W.,  {Middleditch} J.,   {Wang}
  Q.~D.,  1998, \mn@doi [\apjl] {10.1086/311381}, \href
  {https://ui.adsabs.harvard.edu/abs/1998ApJ...499L.179M} {499, L179}

\bibitem[\protect\citeauthoryear{{McConnell}, {McCulloch}, {Hamilton}, {Ables},
  {Hall}, {Jacka}  \& {Hunt}}{{McConnell} et~al.}{1991}]{1991MNRAS.249..654M}
{McConnell} D.,  {McCulloch} P.~M.,  {Hamilton} P.~A.,  {Ables} J.~G.,  {Hall}
  P.~J.,  {Jacka} C.~E.,   {Hunt} A.~J.,  1991, \mn@doi [\mnras]
  {10.1093/mnras/249.4.654}, \href
  {https://ui.adsabs.harvard.edu/abs/1991MNRAS.249..654M} {249, 654}

\bibitem[\protect\citeauthoryear{{Owen} et~al.,}{{Owen}
  et~al.}{2011}]{2011A&A...530A.132O}
{Owen} R.~A.,  et~al., 2011, \mn@doi [\aap] {10.1051/0004-6361/201116586},
  \href {http://adsabs.harvard.edu/abs/2011A%26A...530A.132O} {530, A132}

\bibitem[\protect\citeauthoryear{{Phinney}}{{Phinney}}{1992}]{1992RSPTA.341...39P}
{Phinney} E.~S.,  1992, \mn@doi [Philosophical Transactions of the Royal
  Society of London Series A] {10.1098/rsta.1992.0084}, \href
  {https://ui.adsabs.harvard.edu/abs/1992RSPTA.341...39P} {341, 39}

\bibitem[\protect\citeauthoryear{{Ransom}}{{Ransom}}{2001}]{ransom01}
{Ransom} S.~M.,  2001, PhD thesis, Harvard University

\bibitem[\protect\citeauthoryear{{Ransom}, {Eikenberry}  \&
  {Middleditch}}{{Ransom} et~al.}{2002}]{ransom02}
{Ransom} S.~M.,  {Eikenberry} S.~S.,   {Middleditch} J.,  2002, \mn@doi [\aj]
  {10.1086/342285}, \href
  {https://ui.adsabs.harvard.edu/abs/2002AJ....124.1788R} {124, 1788}

\bibitem[\protect\citeauthoryear{{Scargle}}{{Scargle}}{1982}]{1982ApJ...263..835S}
{Scargle} J.~D.,  1982, \mn@doi [\apj] {10.1086/160554}, \href
  {https://ui.adsabs.harvard.edu/abs/1982ApJ...263..835S} {263, 835}

\bibitem[\protect\citeauthoryear{{Shibata}, {Watanabe}, {Yatsu}, {Enoto}  \&
  {Bamba}}{{Shibata} et~al.}{2016}]{2016ApJ...833...59S}
{Shibata} S.,  {Watanabe} E.,  {Yatsu} Y.,  {Enoto} T.,   {Bamba} A.,  2016,
  \mn@doi [\apj] {10.3847/1538-4357/833/1/59}, \href
  {https://ui.adsabs.harvard.edu/abs/2016ApJ...833...59S} {833, 59}

\bibitem[\protect\citeauthoryear{{Str{\"u}der} et~al.,}{{Str{\"u}der}
  et~al.}{2001}]{2001A&A...365L..18S}
{Str{\"u}der} L.,  et~al., 2001, \mn@doi [\aap] {10.1051/0004-6361:20000066},
  \href {http://adsabs.harvard.edu/abs/2001A%26A...365L..18S} {365, L18}

\bibitem[\protect\citeauthoryear{{Sturm} et~al.,}{{Sturm}
  et~al.}{2013}]{2013A&A...558A...3S}
{Sturm} R.,  et~al., 2013, \mn@doi [\aap] {10.1051/0004-6361/201219935}, \href
  {http://adsabs.harvard.edu/abs/2013A%26A...558A...3S} {558, A3}

\bibitem[\protect\citeauthoryear{{Takata} \& {Cheng}}{{Takata} \&
  {Cheng}}{2017}]{2017ApJ...834....4T}
{Takata} J.,  {Cheng} K.~S.,  2017, \mn@doi [\apj] {10.3847/1538-4357/834/1/4},
  \href {https://ui.adsabs.harvard.edu/abs/2017ApJ...834....4T} {834, 4}

\bibitem[\protect\citeauthoryear{{Titus} et~al.,}{{Titus}
  et~al.}{2019}]{2019MNRAS.487.4332T}
{Titus} N.,  et~al., 2019, \mn@doi [\mnras] {10.1093/mnras/stz1578}, \href
  {https://ui.adsabs.harvard.edu/abs/2019MNRAS.487.4332T} {487, 4332}

\bibitem[\protect\citeauthoryear{{Titus}, {Toonen}, {McBride}, {Stappers},
  {Buckley}  \& {Levin}}{{Titus} et~al.}{2020}]{2020MNRAS.494..500T}
{Titus} N.,  {Toonen} S.,  {McBride} V.~A.,  {Stappers} B.~W.,  {Buckley}
  D.~A.~H.,   {Levin} L.,  2020, \mn@doi [\mnras] {10.1093/mnras/staa662},
  \href {https://ui.adsabs.harvard.edu/abs/2020MNRAS.494..500T} {494, 500}

\bibitem[\protect\citeauthoryear{{Turner} et~al.,}{{Turner}
  et~al.}{2001}]{2001A&A...365L..27T}
{Turner} M.~J.~L.,  et~al., 2001, \mn@doi [\aap] {10.1051/0004-6361:20000087},
  \href {http://adsabs.harvard.edu/abs/2001A%26A...365L..27T} {365, L27}

\bibitem[\protect\citeauthoryear{{Verner}, {Ferland}, {Korista}  \&
  {Yakovlev}}{{Verner} et~al.}{1996}]{1996ApJ...465..487V}
{Verner} D.~A.,  {Ferland} G.~J.,  {Korista} K.~T.,   {Yakovlev} D.~G.,  1996,
  \mn@doi [\apj] {10.1086/177435}, \href
  {https://ui.adsabs.harvard.edu/abs/1996ApJ...465..487V} {465, 487}

\bibitem[\protect\citeauthoryear{{Wang} \& {Zhao}}{{Wang} \&
  {Zhao}}{2004}]{2004ApJ...601.1038W}
{Wang} W.,  {Zhao} Y.,  2004, \mn@doi [\apj] {10.1086/380560}, \href
  {https://ui.adsabs.harvard.edu/abs/2004ApJ...601.1038W} {601, 1038}

\bibitem[\protect\citeauthoryear{{Wilms}, {Allen}  \& {McCray}}{{Wilms}
  et~al.}{2000}]{2000ApJ...542..914W}
{Wilms} J.,  {Allen} A.,   {McCray} R.,  2000, \mn@doi [\apj] {10.1086/317016},
  \href {http://adsabs.harvard.edu/abs/2000ApJ...542..914W} {542, 914}

\bibitem[\protect\citeauthoryear{{Yao}, {Manchester}  \& {Wang}}{{Yao}
  et~al.}{2017}]{yao2017}
{Yao} J.~M.,  {Manchester} R.~N.,   {Wang} N.,  2017, \mn@doi [\apj]
  {10.3847/1538-4357/835/1/29}, \href
  {https://ui.adsabs.harvard.edu/abs/2017ApJ...835...29Y} {835, 29}

\bibitem[\protect\citeauthoryear{{van Straten} \& {Bailes}}{{van Straten} \&
  {Bailes}}{2011}]{vanstraten11}
{van Straten} W.,  {Bailes} M.,  2011, \mn@doi [\pasa] {10.1071/AS10021}, \href
  {https://ui.adsabs.harvard.edu/abs/2011PASA...28....1V} {28, 1}

\bibitem[\protect\citeauthoryear{{van der Heyden}, {Bleeker}  \&
  {Kaastra}}{{van der Heyden} et~al.}{2004}]{2004A&A...421.1031V}
{van der Heyden} K.~J.,  {Bleeker} J.~A.~M.,   {Kaastra} J.~S.,  2004, \mn@doi
  [\aap] {10.1051/0004-6361:20034156}, \href
  {http://adsabs.harvard.edu/abs/2004A%26A...421.1031V} {421, 1031}

\makeatother
\end{thebibliography}






\bsp	
\label{lastpage}

\end{document}